\documentclass[aps,pre,amssymb,floatfix]{revtex4}
\usepackage{bm}
\usepackage{amsmath}
\usepackage{enumerate}
\usepackage{subfigure}
\usepackage{graphicx}
\usepackage{multi row} 

\makeatletter
\renewcommand\subparagraph{\@startsection{subparagraph}{5}{\parindent}%
    {3.25ex \@plus1ex \@minus .2ex}%
    {0.75ex plus 0.1ex}
    {\normalfont\normalsize\bfseries}}
\makeatother
\begin{document}

\title{Nonequilibrium quantum dynamics of many-body systems}
\author{Lea F. Santos}
\affiliation{Department of Physics, Yeshiva University, New York, New York 10016, USA}
\author{ E. Jonathan Torres-Herrera}
\affiliation{ Instituto de F{\'i}sica, Benem\'erita Universidad Aut\'onoma de Puebla, Apt. Postal J-48, Puebla, Puebla, 72570, Mexico}

\begin{abstract}
We review our results for the dynamics of isolated many-body quantum systems described by one-dimensional spin-1/2 models. We explain how the evolution of these systems depends on the initial state and the strength of the perturbation that takes them out of equilibrium; on the Hamiltonian, whether it is integrable or chaotic; and on the onset of multifractal eigenstates that occurs in the vicinity of the transition to a many-body localized phase. We unveil different behaviors at different time scales. We also discuss how information about the spectrum of a many-body quantum system can be extracted by the sole analysis of its time evolution, giving particular attention to the so-called correlation hole. This approach is useful for experiments that routinely study dynamics, but have limited or no direct access to spectroscopy, as experiments with cold atoms and trapped ions.
\end{abstract}
\maketitle

\section{Introduction}
\label{sec:1}
Understanding the properties of interacting many-body quantum systems out of equilibrium is essential to various fields, from atomic, molecular, and condensed matter physics to quantum information and cosmology. Every step forward has the potential to enable new scientific and technological applications. Some examples are listed below.

\begin{enumerate}[(i)]
\item{Nonequilibrium quantum dynamics may reveal new phases of matter that typically do not occur near equilibrium. New phases of matter are tightly connected with the development of new materials, which may revolutionize how we use and produce energy, may lead to new electronic devices, and may give rise to unforeseen innovations.}

\item{Efficient methods to store and transfer many-body quantum coherences are necessary for building analog and digital quantum simulators, developing quantum sensors, and realizing protocols for secure quantum communication. }

\item{One of the most challenging aspects for the development of new electronic devices, such as microchips and hard disks, is the mitigation of local heating. One needs to identify the conditions under which a quantum system can transfer heat rapidly.  }

\item{Further progress in spintronic devices, where information can be transferred without any transport of charge, being therefore better protected against dissipation, requires improved techniques for the control and transport of spin excitations. }
\end{enumerate}

Since the detection of spin echoes in 1950 \cite{Hahn1950}, the goal of studying nonequilibrium quantum dynamics in controllable scenarios became tangible~\cite{Ramanathan2011,WeiARXIV}. A great leap forward came with the Penning and Paul traps, by Hans Dehmelt and Wolfgang Paul, that eventually made possible the study of quantum dynamics of trapped ions~\cite{Leibfried2003,Jurcevic2014,Richerme2014}. Another revolution had as starting point the development and combination of several cooling techniques that culminated in highly controllable experiments with cold atoms, where the evolution of many-body quantum systems is observed for long times~\cite{Bloch2008,Schreiber2015,Kaufman2016}.

The questions that we have been interested in are motivated by those experiments and may also inspire new ones. We have been considering quantum systems with many interacting particles that are well isolated from any sort of environment. By this we mean that the couplings with the surroundings are very weak and can therefore be neglected. Interactions with an environment are unavoidable and do eventually kick in, but we assume that this happens at times much longer than the ones we deal with.

We focus on one-dimensional (1D) many-body quantum systems described by spin-1/2 models. These systems have only two-body interactions and are commonly studied by the experiments mentioned above. The models can also be mapped onto models of spinless fermions and hardcore bosons. In addition, by adjusting the parameters of the Hamiltonian we can cover different regimes (integrable, chaotic, or intermediate between the two), phases (metal vs insulator, ferromagnetic vs antiferromagnetic), symmetries, and strength of the interactions, which allows for the investigation of various different interesting scenarios.

We take the system far from equilibrium by perturbing it faster than any of its characteristic times, so that the perturbation can be seen as instantaneous, what is often referred to as ``quench''. In simple terms, the system is prepared in a nonstationary state. We consider pure states, but the analyses could certainly be extended to mixed states.

Our main goal has been to describe the dynamics of finite lattice many-body quantum systems at different time scales~\cite{Borgonovi2016,HeSantos2013,RigolSantos2010,Santos2008,Rego2009,Santos2009JMP,Santos2011PRE,Santos2012NJP,Santos2012PRL,Santos2012PRE,SantosBernal2015,Santos2016,Bernal2016,Santos2016PRL,Tavora2016,Tavora2017,Torres2014PRA,Torres2014NJP,Torres2014PRAb,Torres2014PRE,TorresKollmar2015,Torres2015,Torres2016BJP,Torres2017,Torres2016Entropy,Torres2017Philo,Torres2017ARXIV,Zangara2013}. Because these systems are finite, their evolution eventually saturates to an equilibrium point. We were able to obtain a detailed description of the so-called survival probability (probability of finding the system still in its initial state later in time) from the moment the system is taken out of equilibrium all the way to the saturation of its evolution~\cite{Tavora2016,Tavora2017,Torres2014PRA,Torres2014NJP,Torres2014PRAb,Torres2014PRE,TorresKollmar2015,Torres2015,Torres2016BJP,Torres2017,Torres2016Entropy,Torres2017Philo,Torres2017ARXIV}. We have not only numerical results, but analytical expressions as well. The survival probability is a simple and insightful quantity. It is part of the equations that compute the temporal evolution of generic physical observables, so having a complete understanding of its evolution provides a better understanding of the dynamics of several observables. This is why we decided to restrict this chapter to this particular quantity and briefly mention others. 

Our main findings for the survival probability, presented in this chapter, are enumerated below.
\begin{enumerate}
\item{The decay of the survival probability depends on the strength of the perturbation. For very strong perturbations, when the energy distribution of the initial state is unimodal, the decay is Gaussian and therefore faster than the usually expected exponential behavior. The Gaussian decay is related to the Gaussian density of states found in many-body systems with two-body interactions~\cite{Torres2014PRA,Torres2014NJP,Torres2014PRE,Torres2014PRAb}.}
\item{Exponential and Gaussian decays are not exclusive to chaotic models and occur also in integrable models perturbed far from equilibrium~\cite{Torres2014PRA,Torres2014NJP,Torres2014PRAb}.}
\item{The speed of the decay depends on the energy of the initial state. The decay is faster for initial states with energy close to the middle of the spectrum, where there is a large concentration of eigenstates, than for states with energies near the border of the spectrum~\cite{Torres2014PRA,Torres2014NJP,Torres2014PRAb}.}
\item{Decays faster than Gaussian occur when the energy distribution of the initial state is bimodal~\cite{Torres2014PRAb}, in which case the quantum speed limit can be reached. Moving away from realistic systems, fast decays can be obtained by increasing the number of particles that interact simultaneously~\cite{Torres2014PRA,Torres2014NJP,Torres2014PRAb}.}
\item{After the initial fast (often Gaussian) decay, the dynamics slows down and becomes power-law. The power-law exponent depends on how the spectrum approaches its energy bounds~\cite{Tavora2016,Tavora2017} and on the level of delocalization of the eigenstates~\cite{Torres2015,Torres2016BJP,Torres2017}.}
\item{In interacting systems with onsite disorder, the value of the power-law decay exponent detects the transition from chaos to many-body localization. This exponent coincides with the fractal dimension of the system~\cite{Torres2015,Torres2016BJP,Torres2017} and with the slope of the logarithmic growth of the Shannon and entanglement entropies~\cite{Torres2017}.}
\item{At long times, after the power-law behavior and before saturation, the survival probability shows a dip below its infinite time average~\cite{Torres2017Philo,Torres2017ARXIV}. This is known as correlation hole and appears only in systems with level repulsion (that is, not in integrable models). The correlation hole provides a way to detect level repulsion from the dynamics, instead of having to resort to the eigenvalues. This is useful for the experiments mentioned above, which have limited access to the spectra of their systems. Since the correlation hole is a general indicator of the integrable-chaos transition, it serves also as a detector of the metal-insulator transition in interacting systems~\cite{Torres2017,Torres2017Philo}.}
\end{enumerate}
Additional highlights of our research, which are not described in this chapter, but may be found in our references, include the following topics.
\begin{enumerate}
\item{The dynamical behavior of the Shannon entropy and entanglement entropy is equivalent~\cite{Torres2016Entropy,Torres2017}. The first is easier to compute numerically and is potentially accessible experimentally, although it is the second that has been mostly studied theoretically.}
\item{Effects associated with the correlation hole are observed also in entropies~\cite{Torres2017Philo} and in experimental observables, such as the spin density imbalance~\cite{Torres2017ARXIV}.}
\item{Analytical expressions for the entire evolution of the survival probability, Shannon entropy, and spin density imbalance can be found using full random matrices~\cite{Torres2017ARXIV,Torres2016Entropy}. Full random matrices are not realistic, but they provide bounds and serve as references to the studies of many-body quantum systems.}
\item{The behavior of the survival probability may signal the presence of an excited state quantum phase transition. It slows down as one approaches the critical point~\cite{Bernal2016,SantosBernal2015,Santos2016}.}
\item{Long-range interactions do not always imply fast dynamics. Depending on the initial state the effects of the long-range couplings may get shielded, resulting in exceedingly slow evolutions~\cite{Santos2016PRL}.}
\item{Despite isolation, one can still talk about equilibration in isolated finite many-body quantum systems, but in a probabilistic sense. By this we mean that after a transient time, few-body observables simply oscillate around their infinite-time average, being very close to it for most time. To speak of equilibration, these temporal fluctuations need to be small and decrease with system size. In Ref.~\cite{Zangara2013}, we show that the size of these fluctuations decrease exponentially with system size in chaotic systems and also in interacting integrable models.}
\item{When the infinite-time averages of few-body observables are very close to microcanonical averages and the difference between the two decreases with system size, we say that the many-body quantum system has thermalized. We have several studies about how the onset of thermalization depends on the initial state and strength of the interactions~\cite{Borgonovi2016,Santos2010PRE,Santos2010PREb,Santos2011PRL,Santos2012PRER,Torres2013,TorresKollmar2015}.}
\end{enumerate}

The text below is divided in two sections. In Sec.~\ref{sec:SpinModel} we provide a pedagogical introduction to the 1D spin-1/2 systems that we study and how to distinguish integrable from chaotic models. In Sec.~3, we present our results for the survival probability for short and long times, from perturbation to saturation.

\section{Spin-1/2 Models}
\label{sec:SpinModel}
We investigate a 1D spin-1/2 system. To describe this chain, one uses spin operators
$S^{x,y,z} = \sigma^{x,y,z}/2$, where 
\[ \sigma^x \equiv \left(
\begin{array}{cc}
0 & 1  \\
1 & 0 
\end{array}
\right) ,
\hspace{0.5 cm}
 \sigma^y \equiv \left(
\begin{array}{cc}
0 & -i  \\
i & 0 
\end{array}
\right) ,
\hspace{0.5 cm}
 \sigma^z \equiv \left(
\begin{array}{cc}
1 & 0  \\
0 & -1 
\end{array}
\right)
\]
are the Pauli matrices and $\hbar$ is set to 1. The quantum state of the spin is represented by a two-component vector (spinor). This state is usually written in terms of the two eigenstates of $S^z$, which then form the basis. One eigenstate represents the spin pointing up in the $z$-direction and the other, the spin pointing down. They can be denoted as
\begin{eqnarray*}
&&|\uparrow \rangle = \left(
\begin{array}{c}
1  \\
0 
\end{array}
\right) , \hspace{2 cm} |\downarrow \rangle = 
\left(
\begin{array}{c}
0  \\
1 
\end{array}
\right) .
\end{eqnarray*}
Since the eigenvalue associated with $|\uparrow \rangle$ is +1/2 and that of  $|\downarrow \rangle$ is -1/2, we refer to the first as the excitation. The operators $S^{x}$ and $S^{y}$ flip the spin up and spin down,
\[
S^{x} |\uparrow \rangle = \frac{1}{2} |\downarrow \rangle \hspace{0.7 cm} S^{x} |\downarrow \rangle = \frac{1}{2} |\uparrow \rangle
\]
\[
S^{y} |\uparrow \rangle = \frac{i}{2} |\downarrow \rangle \hspace{0.7 cm} S^{y} |\downarrow \rangle = - \frac{i}{2} |\uparrow \rangle .
\]

\subparagraph{Basis} 
In a chain with several sites, a commonly used basis in which to write the spin-1/2 Hamiltonian matrix corresponds to arrays where on each site the spin either points up or down in the $z$-direction, as for example $|\downarrow  \uparrow \downarrow \uparrow \downarrow \uparrow \downarrow \uparrow \ldots \rangle_z$. This basis is often referred to as natural-basis, computational-basis or site-basis. We use the latter term.

\subparagraph{Hamiltonian Terms} 
One of the terms that we find in spin-1/2 Hamiltonians is
\begin{equation}
H_Z = \sum_{k} h_k J S_k^z ,
\label{HamOnSite}
\end{equation}
which appears when each site $k$ is subjected to a different local magnetic field. The fields cause the Zeeman splittings of amplitude $h_k J$ on each site. The parameter $J$ sets the energy scale and we choose $J=1$. If all sites have $h_k=h$, that indicates a clean system, where a single magnetic field is applied to the entire chain. If only one site has a Zeeman splitting different from the others, we call it the defect site or the impurity of the system. If all sites have different Zeeman splittings, randomly distributed, then the system is disordered.

When more than one spin is present, they may interact. This may happen through the Ising interaction. If the interaction is active between nearest-neighbors (NN) only, that is sites $k$ and $k+1$, it is given by
\begin{equation}
H_{ZZ} = \sum_{k} J \Delta  S_k^z S_{k+1}^z  ,
\label{HamIsingWithB}
\end{equation}
where $J\Delta$ is the strength of the interaction. This terms causes a pair of adjacent parallel spins to have different energy from a pair of anti-parallel spins, because
\begin{equation}
J \Delta S_k^z S_{k+1}^z |\uparrow_k \uparrow_{k+1}\rangle =+\frac{J\Delta}{4} |\uparrow_k \uparrow_{k+1}\rangle,
\label{upup}
\end{equation}
while
\begin{equation}
J \Delta S_k^z S_{k+1}^z |\uparrow_k \downarrow_{k+1}\rangle =-\frac{J\Delta}{4} |\uparrow_k \downarrow_{k+1}\rangle.
\label{updown}
\end{equation}
The ground state of a Hamiltonian that has only the Ising interaction is ferromagnetic, with all spins aligned in the same direction, when $J \Delta<0$,  and it is antiferromagnetic, with antiparallel neighboring spins, when $J \Delta>0$. We choose the latter.

Another term that appears in our Hamiltonians is the flip-flop term. It interchanges the position of neighboring up- and down-spins according to 
\[
J(S_k^x S_{k+1}^x + S_k^y S_{k+1}^y)|\uparrow_k \downarrow_{k+1} \rangle =
\frac{J}{2} |\downarrow_k \uparrow_{k+1} \rangle .
\]
The NN flip-flop term couples site-basis vectors that differ only by the orientation of the spins in two neighboring sites. When the Hamiltonian matrix is written in the site-basis, the flip-flop term constitutes the off-diagonal elements.

\subparagraph{Spin-1/2 Hamiltonian} 
The Hamiltonian that we consider is a combination of the terms described above. It is given by
\begin{eqnarray}
H &=& d J S_{L/2}^z + \sum_{k=1}^{L} h_k J S_k^z \label{ham} \\
&+& J\sum_{k} \left( S_k^x S_{k+1}^x + S_k^y S_{k+1}^y +\Delta S_k^z S_{k+1}^z \right) + \lambda J\sum_{k} \left( S_k^x S_{k+2}^x + S_k^y S_{k+2}^y +\Delta S_k^z S_{k+2}^z \right). \nonumber 
\end{eqnarray}
The chain has $L$ sites and we denote by $N_{up}$ the number of up-spins. The amplitude $dJ$ indicates the Zeeman splitting of the defect site.  The Zeeman splittings $h_k J$ correspond to onsite disorder caused by random static magnetic fields; $h_k$ are random numbers from a uniform distribution in $[-h,h]$ and $h$ is the strength of the disorder.  $\Delta$ is the anisotropy parameter; when the Ising interaction and the flip-flop term have the same strength ($\Delta=1$), the system is isotropic. $\lambda$ is the ratio between the NN and next-nearest-neighbor (NNN) couplings.  

Depending on the boundary conditions, we refer to the chain as open or closed. Open boundary conditions imply that a spin on site 1 can only couple with a spin on site 2 and a spin on site $L$ can only couple with a spin on site $L-1$. In closed (or periodic) boundary conditions the chain is a ring, where a spin on site 1 can couple with a spin on site 2 and also with a spin on site $L$. The index in the second and third sums of Eq.~(\ref{ham}) runs according to the boundaries.

\subsection{Symmetries}
Any symmetry of the system is associated with an operator $O$ that commutes with the Hamiltonian. According to Noether's theorem, this operator represents a constant of motion, as seen from 
$ \dfrac{d O}{dt} =i[H,O]$. For the Hamiltonian in Eq.~(\ref{ham}), we identify the following symmetries.

\begin{enumerate}
\item{ $H$ commutes with the total spin in the $z$-direction, ${\cal S}^z=\sum_{k=1}^L S_k^z$. The system conserves ${\cal S}^z$; it is invariant by a rotation around the $z$-axis. This means that the eigenstates of $H$ are also eigenstates of ${\cal S}^z$, so they have a fixed number of spins pointing up. Each eigenstate $|\psi\rangle$ is a superposition that involves only site-basis vectors with the same number of up-spins. For example, for $L=4$ and ${\cal S}^z=0$ we have,
\[
|\psi \rangle = C_1 |1100\rangle +  C_2 |1010\rangle + C_3  |1001\rangle + C_4  |0110\rangle + C_5  |0101\rangle + C_6  |0011\rangle,
\]
where $C_n$'s are the probability amplitudes, $n=1, \ldots {\cal D}$, and ${\cal D}$ is the dimension of the subspace. 
The Hamiltonian matrix of a system with $L$ sites written in the site-basis is composed of $L+1$ independent blocks (or subspaces), each with a fixed number of up-spins, $N \in [0,L]$. The dimension of each block is ${\cal D}=L!/[(L-N)!N!]$.}

\item{When $d, h=0$, Hamiltonian (\ref{ham}) is invariant under reflection, which leads to conservation of parity, that is, $H$ commutes with the parity operator 
\[
\Pi = \left\{
\begin{array}{ccc}
{\cal P}_{1,L} {\cal P}_{2,L-1} \ldots {\cal P}_{\frac{L}{2},\frac{L+2}{2}} & {\rm for} & {\rm L=even}\\
{\cal P}_{1,L}  {\cal P}_{2,L-1} \ldots {\cal P}_{\frac{L-1}{2},\frac{L+3}{2}} & {\rm for} & {\rm L=odd}
\end{array}
\right.
\]
where ${\cal P}_{k,l}=(\sigma^x_k \sigma^x_l + \sigma^y_k \sigma^y_l + \sigma^z_k \sigma^z_l +\mathbb{1})/2$ is the permutation operator and $\mathbb{1}$ is the identity operator.  ${\cal P}_{k,l}$ swaps the states of the $k^{th}$ and $l^{th}$ sites. For example, for $L=4$ and a single excitation, $N_{up}=1$, the probability amplitudes  in
$|\psi \rangle = a_{1} |\uparrow \downarrow \downarrow \downarrow \rangle +
a_{2} |\downarrow \uparrow \downarrow \downarrow \rangle +
a_{3} |\downarrow \downarrow \uparrow \downarrow \rangle +
a_{4} |\downarrow \downarrow \downarrow \uparrow \rangle 
$ are either $a_{1}=a_{4}$ and $a_{2}=a_{3} $ for even 
parity or $a_{1}=- a_{4}$ and $a_{2}=- a_{3} $ for odd 
parity.}

\item{When $d, h=0$, $L$ is even, and $N_{up}=L/2$, Hamiltonian (\ref{ham}) is invariant under a global $\pi$ rotation around the $x$-axis. The operator that represents the rotation is
\[
R^x_{\pi} = \sigma^x_1  \sigma^x_2 \ldots \sigma^x_L
\] 
As an example, take $L=4$ and $N_{up}=2$. The eigenstate 
\[
|\psi \rangle = a_{1}|\uparrow \uparrow \downarrow \downarrow  \rangle +
a_{2} |\uparrow \downarrow \uparrow  \downarrow \rangle +
a_{3} |\uparrow \downarrow \downarrow \uparrow  \rangle 
+
a_{4} |\downarrow \uparrow \uparrow \downarrow  \rangle +
a_{5} |\downarrow \uparrow  \downarrow  \uparrow \rangle +
a_{6} |\downarrow \downarrow \uparrow \uparrow  \rangle 
\]
has either $a_{1}=a_{6}$, $a_{2}=a_{5}$, and $a_{3}=a_{4}$ or $a_{1}=- a_{6}$, $a_{2}=- a_{5}$, and $a_{3}=- a_{4}$.}

\item{When $d, h=0$ and $\Delta=1$, the total spin ${\cal S}_T = \sum_n \vec{S}_n$ is conserved.}

\end{enumerate}
We can break the symmetries listed above, except for the total spin in the $z$-direction, as follows. Conservation of total spin can be avoided by choosing $\Delta \neq 1$. Parity and spin reversal can be broken if we deal with an open chain and add an impurity on a site in the border of the chain.

\subsection{Integrable vs chaotic models}
In classical mechanics, if a system with $n$ degrees of freedom has $n$ independent integrals of motion that are Poisson-commuting, then the system is integrable. In this case, the differential equations describing the time evolution can be explicitly integrated using action-angle variables. The solutions display periodic motion on tori in phase space, and ergodicity is nonexistent. In contrast to the classical case, the notion of integrability at the quantum level has been a source of debates~\cite{SutherlandBook,Caux2011}.

In the case of Hamiltonian (\ref{ham}), we use the term integrability when referring to choices of parameters that allows the Hamiltonian to be solved with the Jordan-Wigner transformation or the Bethe Ansatz~\cite{Bethe1931}. We select the following two cases,  respectively.
\begin{enumerate}
\item{The XX model is a noninteracting integrable model, where $d, h, \Delta, \lambda=0$. }
\item{The XXZ model is an integrable interacting model, where $\Delta \neq 0$ and $d, h,\lambda=0$. }
\end{enumerate}

The notion of quantum chaos is another delicate subject. Classical chaos goes back to the studies of Poincar\'e. It is related to the extreme sensitivity of the dynamics of a system to its initial conditions. The main features of classical chaos can be illustrated with a dynamical billiard. It corresponds to an idealized billiard table that has no friction and where a particle reflects elastically from the boundaries. The motion of the particle is represented in phase space by a trajectory  restricted to a surface of constant energy. The shape of the boundaries determines whether the system is chaotic or regular. In the first case,  two trajectories with very close initial conditions diverge exponentially in time. The rate of this separation is  the Lyapunov exponent. The trajectories may become ergodic, in which case, after a long time, the particle will have visited the entire surface of constant energy and will be equally likely to be found in any point of the accessible phase space.

For quantum systems the notion of phase-space trajectories loses its meaning, since as stated by the Heisenberg uncertainty principle, we can no longer precise the particle's position and momentum at the same time. However, since classical physics is a limit of quantum physics, we could still search for quantum signatures of classical chaos.

The term quantum chaos refers to properties of eigenvalues and eigenstates that are found in the quantum level and indicate whether the system in the classical level is chaotic. It has been conjectured that the spectral fluctuations in the quantum limit of classical system that is chaotic are always the same~\cite{Casati1980,Bohigas1984}. This conjecture has been proved in the semiclassical limit. The term has also been extended to include quantum systems without a classical limit, as our spin-1/2 models. 

The distribution of the spacings between neighboring energy levels of a quantum system is the most commonly employed tool to distinguish integrable from nonintegrable models, but others exist, such as the level number variance and the spectral rigidity~\cite{Guhr1998}, as well as the distribution of the ratio of consecutive level spacings~\cite{Atas2012}. If the system is chaotic, the energy levels are highly correlated and repel each other; if it is regular (integrable), the energy levels are uncorrelated, randomly distributed, and can cross.  The chaotic spin models associated with Hamiltonian~(\ref{ham}) include:

\begin{enumerate}
\item{The defect model with no random disorder ($h=0$), $0<\Delta<1$, $0<d<1$, and $\lambda=0$. The interplay between the defect and the impurity drives the system into the chaotic domain~\cite{Santos2004,Torres2014PRE}.}
\item{The NNN model with no random disorder ($h=0$) and $0<\lambda < 1$ \cite{Hsu1993,Kudo2005,Santos2009JMP,Gubin2012}. The system remains chaotic whether $\Delta \neq 0$ or $\Delta=0$ \cite{Santos2011PRE}.}
\item{The disordered model with $0<h<1$, $\Delta=1$, $d, \lambda =0$ \cite{Santos2004,SantosEscobar2004,Torres2015,Torres2017}.}
\end{enumerate}

\subsubsection{Unfolding procedure}
When studying the level spacing distribution, to be able to compare different systems of different sizes, and also the different parts of the spectrum of the same system, we need to unfold the spectrum. This means that each system's specific mean level density must be removed from the data. It does not make sense to compare local fluctuations from systems with very different average densities. For example, it does not make sense to say that a spectral region with high average density has less repulsion than a spectral region with low average density. We need to separate the local fluctuations from a systematic global energy dependence of the average density. For this, we rescale the energies, so that the mean level spacing is 1. Since the density of states is the number of states in an interval of energy, that is, the reciprocal of the mean level spacing, this renormalization procedure ensures also that the mean local density of states becomes unit.

There are different ways to unfold the spectrum.  A simple and good enough recipe is the following~\cite{Gubin2012}.

(i) Order the spectrum in increasing values of energy.

(ii) Discard some eigenvalues from the edges of the spectrum, where the fluctuations are large. This is arbitrary, we can discard for example 10\% of the eigenvalues.

(ii) Separate the remaining eigenvalues into small sets of eigenvalues.

(iii) For each set, divide the eigenvalues by the mean level spacing of that particular set. The mean level spacing of the new set of renormalized energies is now 1.

Notice that contrary to the level spacing distribution, the distribution of the ratio of consecutive level spacings does not require the unfolding of the spectrum~\cite{Atas2012}.

\subsubsection{Level spacing distribution of integrable models}
\begin{figure}[ht]
\includegraphics*[scale=.33]{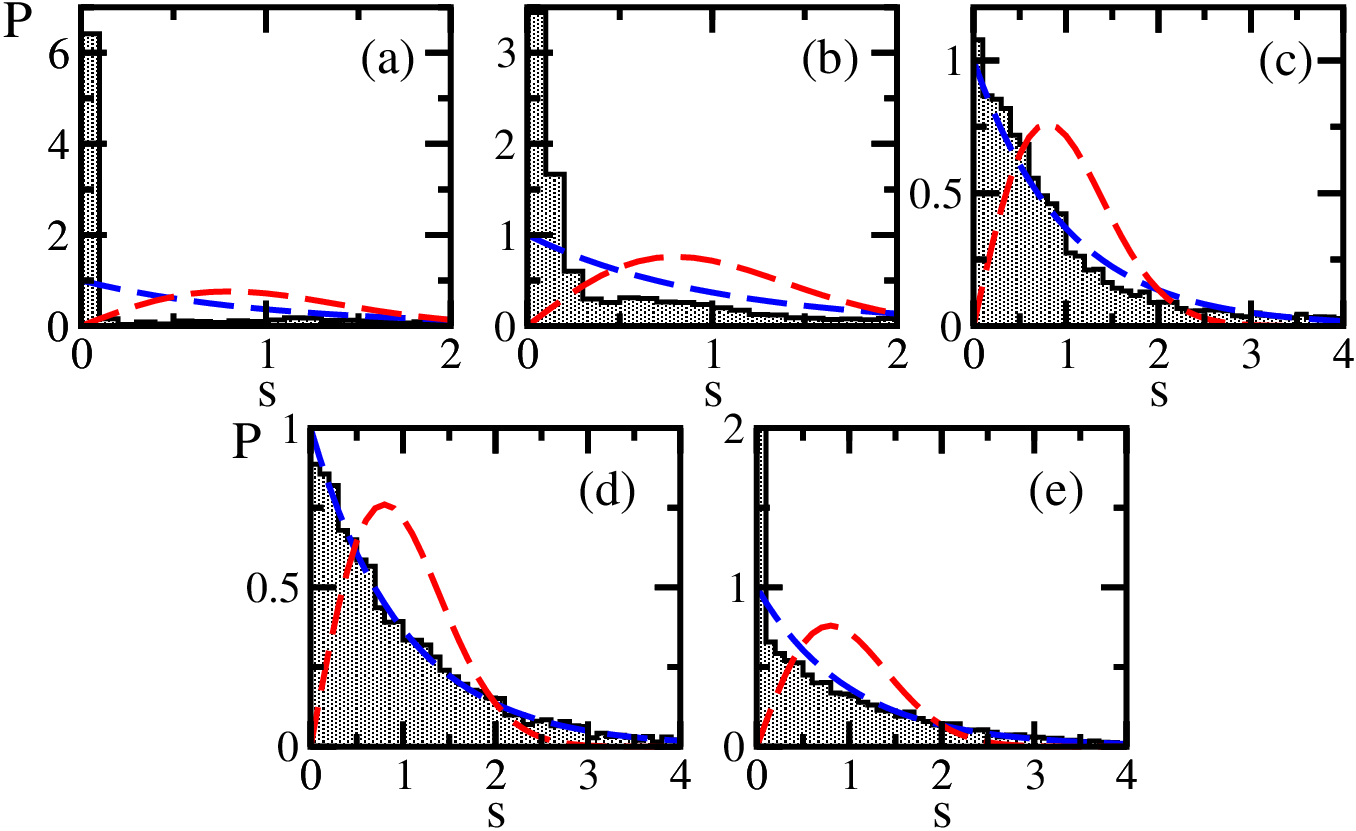}
\caption{Level spacing distribution for a single subspace and eigenstates with even parity; $L=18$, $N_{up}=6$, $d, h, \lambda=0$, open boundary conditions. The Poisson and Wigner-Dyson distributions are shown with dashed lines. From (a) to (e): $\Delta=0.0, 10^{-3}, 10^{-2}, 0.1, 0.5$.}
\label{Fig:Poisson}     
\end{figure}

In integrable models, the eigenvalues are uncorrelated, they are not prohibited from crossing and usually follow Poisson statistics. The distribution $P(s)$ of the neighboring spacings $s$ is given by
\begin{equation}
P_{\rm P}(s) = e^{-s}.
\end{equation} 
However, deviations from this shape are seen for the XX model due to its the high number of degeneracies. As $\Delta$ increases from zero, the excessive degeneracies rapidly fade away and the Poisson distribution is recovered [compare Figs.~\ref{Fig:Poisson} (a) and (b) with Figs.~\ref{Fig:Poisson} (c) and (d)]. At the root of unit $\Delta=1/2$, the distribution departs again from Poisson (see Figs.~\ref{Fig:Poisson} (e)].  By changing $\Delta$ slightly, for example, by using = 0.48, the Poisson distribution reappears \cite{Zangara2013}.

\subsubsection{Level spacing distribution of chaotic models}
The level spacing distribution $P(s)$  of chaotic models is given by the Wigner-Dyson (WD) distribution~\cite{MehtaBook,HaakeBook,Guhr1998,ReichlBook}. The specific form of the Wigner-Dyson distribution depends on the symmetries of the Hamiltonian. In time-reversal invariant systems with rotational symmetry, the Hamiltonian is represented by real and symmetric matrices,
$H_{ij} = H_{ji}$, and the level spacing distribution has the following shape,
\begin{equation} 
P_{\rm WD}(s) = \frac{\pi }{2} s \exp \left( {-\frac{\pi}{4} s^2} \right),
\label{Ps_GOE}
\end{equation}
which makes evident the level repulsion. This expression was derived exactly for 2$\times$2 matrices and it works extremely well for large matrices~\cite{MehtaBook,HaakeBook}.

\begin{figure}[ht]
\includegraphics*[scale=.43]{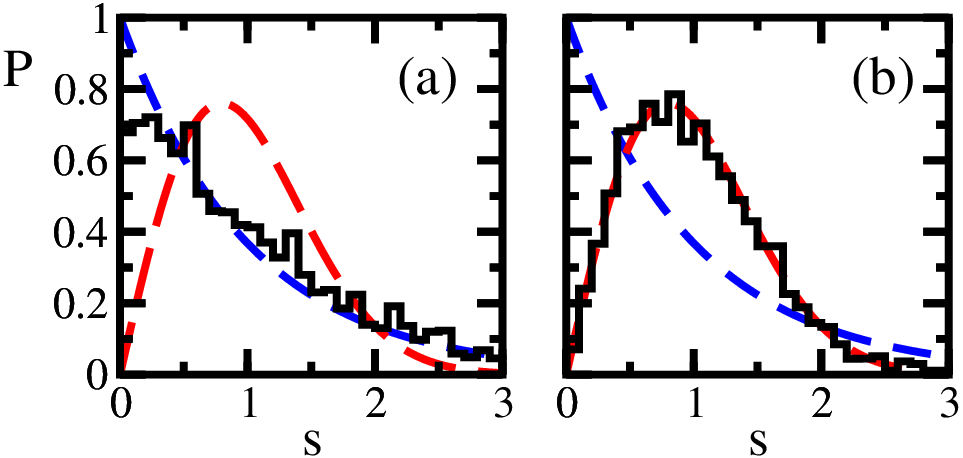}
\caption{Level spacing distribution for the chaotic Hamiltonians written in the text.
Panel (a): $L=14$, $N_{up}=7$, ${\cal S}^z=0$.
Panel (b): $L=15$, $N_{up}= 5$ and 
the eigenvalues are separated by 
the parity;
$P(s)$ is the average of the distributions 
of the two parity sectors. The Poisson and Wigner-Dyson distributions are shown with dashed lines.}
\label{Fig:WD}     
\end{figure}

To obtain the level spacing distribution, we need to separate the eigenvalues according to their symmetry sectors. If we mix eigenvalues from different symmetry sectors, we may not achieve a Wigner-Dyson distribution even if the system is chaotic, because eigenvalues from different subspaces are independent, uncorrelated, are have no reason to repel each other~\cite{Santos2009JMP,Gubin2012}. To illustrate this issue, in Fig.~\ref{Fig:WD} we consider two chaotic Hamiltonians with open boundary conditions. They represent clean systems, where $d, h=0$. The Hamiltonians are chaotic and given by:

\vskip 0.4 cm
(a) 
$H = \sum_{k=1}^{L-1} \left[ \left(
S_k^x S_{k+1}^x + S_k^y S_{k+1}^y \right) +
 S_k^z S_{k+1}^z \right] $
 
 $\hspace{0.9 cm} + 0.5 \sum_{k=1}^{L-2} \left[  \left(
S_k^x S_{k+2}^x + S_k^y S_{k+2}^y \right) +
 S_k^z S_{k+2}^z \right].$

\vskip 0.4 cm
(b) 
$H = \sum_{k=1}^{L-1} \left[ \left(
S_k^x S_{k+1}^x + S_k^y S_{k+1}^y \right) +
0.5 S_k^z S_{k+1}^z \right] $

$\hspace{0.9 cm} + 0.5 \sum_{k=1}^{L-2} \left[  \left(
S_k^x S_{k+2}^x + S_k^y S_{k+2}^y \right) +
0.5 S_k^z S_{k+2}^z \right].$

\vskip 0.4 cm
In both panels of Fig.~\ref{Fig:WD} we have eigenvalues of a single selected ${\cal S}^z$-sector. In panel (b),  we avoid the ${\cal S}^z=0$ subspace, where spin reversal symmetry exists, by choosing $L$ odd.  We also choose $\Delta \neq 1$ to avoid conservation of total spin. In doing so, the only remaining symmetry is parity, which we do take into account. The expected  Wigner-Dyson distribution is found. Contrary to Fig.~\ref{Fig:WD} (b), Fig.~\ref{Fig:WD} (a) mixes eigenvalues from the three symmetries mentioned above -- spin reversal, total spin, and parity -- which explains why $P(s)$ becomes so close to a misleading Poisson distribution.

\subsubsection{Level spacing indicator}

To study the crossover from integrability to chaos as a certain parameter is varied, better than plotting the level spacing for each value of the parameter, we can use a quantity that tells us how close we are to a Poisson or to a Wigner-Dyson distribution. An example is the indicator $\beta$ used to fit $P(s)$ with the Brody distribution~\cite{Brody1981},
\begin{equation}
P_B(s) = (\beta +1) b s^{\beta} \exp \left( -b s^{\beta +1} \right), \hspace{0.2 cm}
b= \left[\Gamma \left( \frac{\beta + 2}{\beta +1} \right)\right]^{\beta +1},
\label{eq:BrodyParameter}
\end{equation}
where $\Gamma$ is Euler's gamma function.
If $\beta=0$ the distribution is Poisson and $\beta=1$ indicates a Wigner-Dyson distribution. 

Based on heuristic arguments,  Izrailev introduced an Ansatz for the level spacing distribution that captures very well the intermediate regime between Poisson and Wigner-Dyson \cite{Izrailev198813,Izrailev1990}, 
\begin{equation}\label{eq:IzrailevDistro}
P_I(s)=A s^{\nu} \exp \left[-\frac{\pi^2}{16}\nu s^2-(C-\frac{\nu}{2})\frac{\pi}{2}s\right]\,,
\end{equation}
where $A$ and $C$ are constants with values obtained through normalization conditions. The parameter $\nu$ in Eq.~(\ref{eq:IzrailevDistro}) is related to the degree of localization of the eigenstates. For $\nu=0$, we can recover the Poisson distribution. For $\nu=1$, we have the GOE level repulsion, that is $P(s\to0) \to s$. 

Another way to quantify the proximity to the Wigner-Dyson distribution is with the chaos indicator~\cite{Jacquod1997}
\begin{equation}
\eta = \frac{\int_0^{s_0} [P(s) - P_{WD}(s)] ds}{\int_0^{s_0} [P_P(s) - P_{WD}(s)] ds}, 
\end{equation}
where $s_0$ is the first intersection point of $P_P(s)$ and $P_{WD}(s)$. For a Poisson distribution, $\eta \rightarrow 1$, and for the Wigner-Dyson, $\eta \rightarrow 0$.

\begin{figure}[ht]
\includegraphics*[scale=.4]{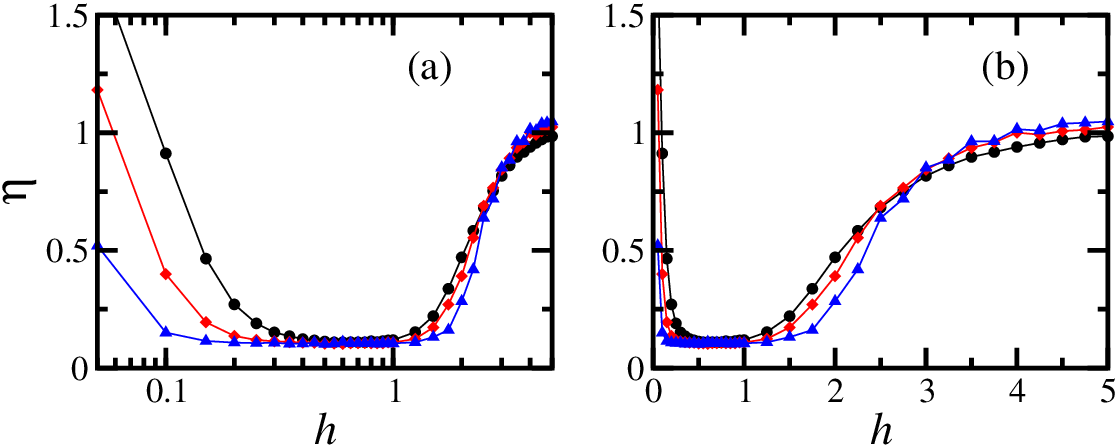}
\caption{Chaos indicator $\eta$ {\em vs.} disorder strength; semilogarithmic scale (a) and linear scale (b). The system sizes are $L = 12$ (circles), $L = 14$ (diamonds), and $L = 16$ (triangles); $\Delta=1$, $d, \lambda=0$, $N_{up}=L/2$. Average performed over $1082, 291, 77$ disorder realizations for $L=12,14,16$, respectively.}
\label{Fig:indicator}     
\end{figure}

\subparagraph{Disordered Spin Model} 
 In Fig.~\ref{Fig:indicator}, we show $\eta$ as a function of $h$ for the disordered model with $\lambda=0$, $\Delta=1$, and closed boundary conditions. $\eta$ is averaged over several disorder realizations.  As the disorder strength $h$ increases from zero (where we have the clean integrable XXZ model), the level spacing distribution first transitions abruptly from Poisson ($\eta \sim 1$) to Wigner-Dyson (small $\eta$).  For the system sizes considered, it remains Wigner-Dyson for $h$ in the range $[0.1,1]$, where $\eta$ plateaus to a small value. As $h$ further increases, the level spacing distribution transitions from Wigner-Dyson back to Poisson. In this second integrable region, the system becomes localized in space~\cite{Santos2004,SantosEscobar2004,Dukesz2009}.

The logarithmic scale of the $x$-axis in Figs.~\ref{Fig:indicator} (a)  emphasizes the first transition from the spatially delocalized integrable point to chaos and the linear plot in Figs.~\ref{Fig:indicator} (b) stresses the transition to localization in space. The different curves represent different system sizes; they increase from top to bottom. The range of disorder strengths for which $\eta $ is small increases as $L$ increases. This indicates that in the thermodynamic limit, the two transition regions may disappear, although this is still an open question. An infinitesimally small $h$ may suffice to take the system into the chaotic regime~\cite{Santos2010PRE,Torres2014PRE}. As  for $h>1$, the transition region may disappear in the thermodynamic limit or persist, in this latter case, maybe as a critical point as one finds in Anderson localization in higher dimensions.

\subsection{Density of States and Delocalization Measures}
\label{subsec:DOS}
The Wigner-Dyson distribution was first studied in the context of full random matrices. Wigner \cite{Wigner1951} employed these matrices to describe the spectrum of heavy nuclei.  His idea was to ignore the details of the interactions of such complex systems and treat them statistically. Full random matrices are filled with random numbers and their only constraint is to satisfy the symmetries of the system one is trying to describe. In the case of Gaussian orthogonal ensembles (GOE), the full random matrices are invariant with respect to an orthogonal conjugation $O^{T}HO$, where $O$ is any orthogonal matrix (that is a matrix whose inverse is equal to its transpose). GOE random matrices are real and symmetric, as the Hamiltonian matrices for the spin systems that we study. 

The level spacing distribution of GOE random matrices is also given by Eq.~(\ref{Ps_GOE}) and agrees extremely well with the distributions obtained with data from actual nuclei spectra. However, full random matrices are unrealistic, since they assume simultaneous and infinite-range interactions among all the particles of the system that they try to represent. In contrast, our spin models  describe realistic systems studied experimentally. They have only two-body and short-range interactions. What are then the properties that clearly distinguish realistic models with two-body interactions from full random matrices? 

\begin{figure}[ht]
\includegraphics*[scale=.42]{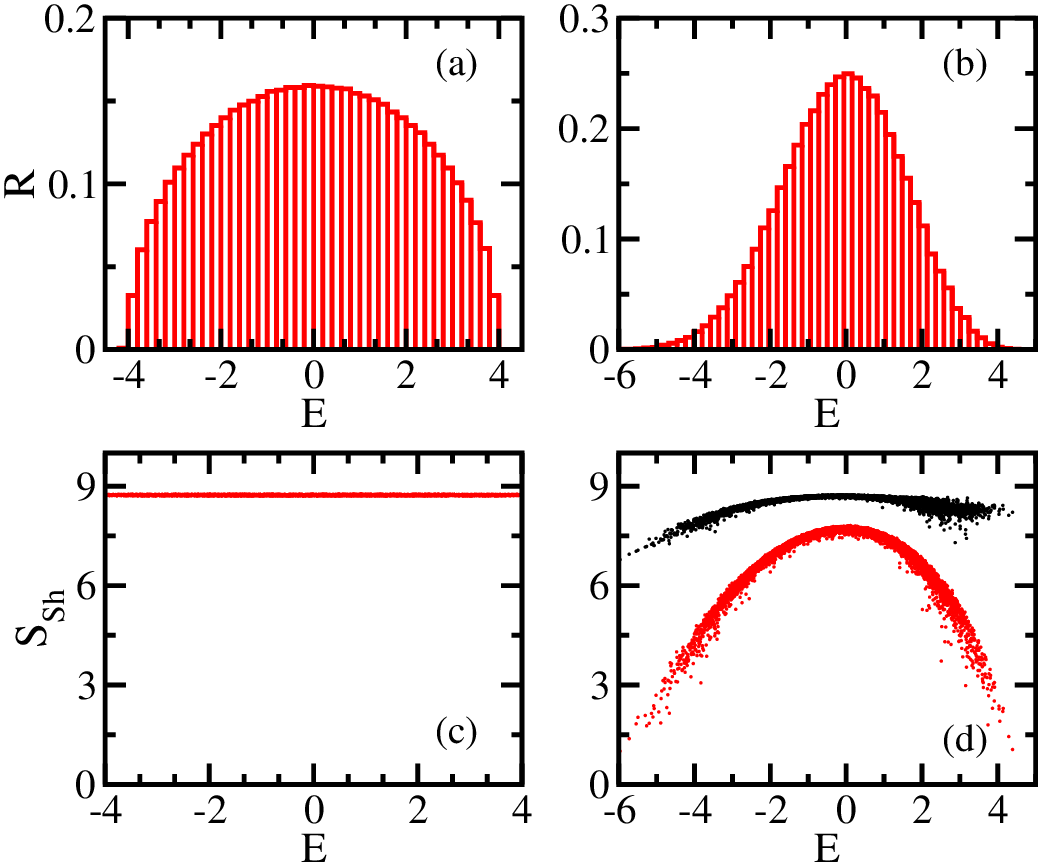}
\caption{Left: GOE full random matrix with ${\cal D}=12870$. Right: Defect model with  $h, \lambda=0$,  $\Delta=0.5$, $d=0.9$,  $L=16$, $N_{up}=8$, ${\cal D}=12870$, open boundary conditions. The random numbers of the full random matrix are rescaled so that ${\cal E} = 4$. Top: Density of states. Bottom: Shannon entropy for all eigenstates. In (d): site-basis (black) and mean-field basis (red).}
\label{Fig:DOS}     
\end{figure}

A distinctive feature between full random matrices and realistic models is the density of states,
\begin{equation}
R(E) = \sum_{\alpha}  \delta (E - E_\alpha ),
\end{equation}
where $E_{\alpha}$ are the eigenvalues of the Hamiltonian. The density of states of full random matrices follows the standard semicircle distribution~\cite{Wigner1955},
\begin{equation}
R (E)= \frac{2}{\pi {\cal E}} \sqrt{1- \left( \frac{E}{{\cal E}} \right)^2 },
\label{DOSfrm}
\end{equation}
where $2 {\cal E}$ is the length of the spectrum, that is $-{\cal E} \leq E \leq {\cal E}$.  The density of states of  Hamiltonians with two-body interactions  is Gaussian, independent of the regime (integrable or chaotic) of the system. These two cases are illustrated in Figs.~\ref{Fig:DOS} (a) and (b) for full random matrices and the defect model, respectively.

The Gaussian shape of the density of states of realistic models is reflected into the structure of the eigenstates. The majority of the eigenstates are close to the middle of the spectrum, where strong mixing can then take place and the eigenstates reach their highest level of delocalization. In contrast, the eigenstates closer to the edges of the spectrum are more localized.

There are various ways to quantify how much a state spreads out in a certain basis. One of them is the participation ratio $PR$. Given an eigenstate $|\psi_{\alpha} \rangle =  \sum_n C_n^{\alpha} |\phi_n\rangle$ written in a basis $|\phi_n\rangle$,  
\begin{equation}
PR^{(\alpha)} = \frac{1}{\sum_n |C_n^{\alpha}|^4}.
\end{equation}
 A comparable quantity is the Shannon information entropy, defined as
\begin{equation}
S_{Sh}^{(\alpha)} = - \sum_n |C_n^{\alpha}|^2 \ln |C_n^{\alpha}|^2.
\end{equation}
The values of $PR^{(\alpha)}$ and $S_{Sh}^{(\alpha)}$ depend on the chosen basis. In the case of full random matrices,  the notion of basis is not well defined. All eigenstates of full random matrices are (pseudo)-random vectors. In the case of GOEs, the coefficients are real random numbers from a Gaussian distribution satisfying the normalization condition. All eigenstates are therefore equivalent and lead to approximately the same values of the participation ratio and of the Shannon entropy~\cite{Torres2016Entropy},
\begin{equation}
PR^{GOE}\sim {\cal D}/3, \hspace{0.9 cm} S_{Sh}^{GOE} \sim \ln (0.48 {\cal D}). 
\end{equation}

The results above can be obtained by substituting the sum in $PR$ and $S_{Sh}$ by an integral,
\[
\sum_n F(C_n ) \rightarrow {\cal D} \int_{-\infty}^{\infty} F(C) P(C) dC.
\]
 The distribution of the probability amplitudes $C^{\alpha}_{n} $ is given by the Gaussian~\cite{ZelevinskyRep1996},
\[
P(C) = \sqrt{  \frac{{\cal D}}{2 \pi}   }  \exp \left( - \frac{{\cal D}}{2 } C^2 \right),
\]
so $\overline{C}=0$ and $\overline{C^2}=1/{\cal D}$. The latter is obtained by substituting $x = C \sqrt{{\cal D}/2} $,
\[
\overline{C^2} = \sqrt{  \frac{{\cal D}}{2 \pi}   }  \int_{-\infty}^{\infty}  dC C^2\exp \left( - \frac{{\cal D}}{2 } C^2 \right) = 
\sqrt{  \frac{{\cal D}}{2 \pi}   }  \int_{-\infty}^{\infty} \frac{2}{{\cal D}} dx x^2 e^{-x^2} \sqrt{\frac{2}{{\cal D}}} = \frac{1}{{\cal D}}.
\]
Thus, for the inverse of the participation ratio, we have
\begin{eqnarray}
\sum_n |C_n|^4  \rightarrow {\cal D} \sqrt{  \frac{{\cal D}}{2 \pi}   }  \int_{-\infty}^{\infty} dC C^4 \exp \left( - \frac{{\cal D}}{2 } C^2 \right) &=&
 {\cal D} \sqrt{  \frac{{\cal D}}{2 \pi}   }  \int_{-\infty}^{\infty} dx \frac{4}{{\cal D}^2} x^4 e^{-x^2} \sqrt{\frac{2}{{\cal D}}} 
 \nonumber \\
 &=& 
\frac{4}{{\cal D}\sqrt{\pi} }    \frac{3 \sqrt{\pi}}{4} = \frac{3}{{\cal D}}  \nonumber
\end{eqnarray}
and for the Shannon information entropy,
\[
S_{Sh}^{GOE} \sim - {\cal D} \sqrt{\frac{{\cal D}}{2 \pi}} \int_{-\infty}^{\infty} \exp \left( - \frac{{\cal D} C^2}{2}  \right) C^2 \ln C^2  dC= 
-2 +  \ln 2 + \gamma_e  + \ln{\cal D}  \sim \ln (0.48 {\cal D}) ,
\]
where $\gamma_e$ is Euler's constant.

In Figs.~\ref{Fig:DOS} (c) and (d), we show the Shannon entropy for full random matrices and the defect model, respectively. For the first, apart from small fluctuations, $S_{Sh} = S_{Sh}^{GOE}$. For the realistic model, we show $S_{Sh}$ for the eigenstates written in two different basis representation. The choice of basis depends on the problem we are interested in. For studies of localization in real space, the site-basis  is a natural choice. Another alternative, often used to distinguish the regular from the chaotic region, is the mean-field basis, which corresponds to the integrable (regular) part of the Hamiltonian. In the case of the defect model, a reasonable choice for the mean-field basis corresponds to the eigenstates of the XXZ model~\cite{Santos2012PRL,Santos2012PRE}. Both cases are shown in Fig.~\ref{Fig:DOS} (d).

The energy dependence of the structure of the eigenstates of a realistic system has consequence for its dynamics~\cite{Santos2012PRL,Santos2012PRE,Torres2014PRA,Torres2014NJP,Torres2014PRAb,TorresKollmar2015,TorresProceed,Torres2016BJP,Torres2016Entropy,Tavora2016,Tavora2017} and viability of thermalization~\cite{Santos2010PRE,Santos2010PREb,RigolSantos2010,Torres2013,Borgonovi2016}. The dynamics is slower for an initial state with energy close to the edge of the spectrum than for an initial state close to the middle of the spectrum. Thermalization is expected for chaotic systems, but it may not occur for initial states with energies very close to the border of the spectrum.

\section{Dynamics: Survival Probability}
\label{sec:SurvProb}
Now that we have a general idea about the spectrum and the structure of the eigenstates of systems with two-body interactions, as those described by spin-1/2 models [Eq.~(\ref{ham})], we proceed with the analysis of their dynamics. We assume that the system is prepared  in an initial state $|\Psi (0)\rangle$ that is an eigenstate of a certain initial Hamiltonian $H_0$. The dynamics starts with the sudden change (quench) of a parameter of the Hamiltonian that brings it to a new final Hamiltonian $H$,
\begin{equation}
H_0 \xrightarrow{quench} H=H_0+gV,
\label{eq:quench}
\end{equation}
where $g$ is the strength of the perturbation. 

There are various quantities that we can use to analyze the evolution of the system. We look here at the simplest one: the probability of finding the system at time $t$ still in state $|\Psi (0)\rangle$, which is known as the survival probability and is given by
\begin{equation}
W_{n_0}(t)=|\langle \Psi (0) |\Psi (t)\rangle |^2=|\langle \Psi (0)| e^{ - iHt} |\Psi (0)\rangle |^2 .
\label{eq:F}
\end{equation}
$W_{n_0}(t)$ is also known as nondecay probability, return probability, or fidelity, but it is incorrect to call it Loschmidt echo, since we have only evolution forward,  there is no time reversal (``echo'')  involved. 
 
By writing the initial state in the eigenstates $|\psi_{\alpha}\rangle$ of $H$, Eq.~(\ref{eq:F}) becomes
\begin{eqnarray}
W_{n_0}(t) = \left| \sum\limits_\alpha  | C^\alpha _{n_0} |^2 e^{ - i E_\alpha t } \right|^2 =  \left| \int dE\, e^{ - iEt} \rho_0(E) \right|^2,
\label{saitorho}
\end{eqnarray} 
where $C^{\alpha}_{n_0} =\langle \psi_{\alpha} | \Psi(0) \rangle$ are the overlaps and
\begin{equation}
\rho_0(E)  \equiv \sum_{\alpha}  | C^{\alpha}_{n_0} |^2 \delta (E - E_\alpha )
\end{equation} 
is the energy distribution of the initial state, referred to as local density of states (LDOS). The survival probability is the absolute square of the Fourier transform of the LDOS. If we have detailed information about $\rho_0(E)$ we can predict the behavior of $W_{n_0}(t)$.
The mean and variance of the LDOS are respectively the energy of the initial state,
\begin{equation}
E_0= \langle \Psi (0)| H |\Psi (0)\rangle = \sum_{\alpha} |C^{\alpha}_{n_0}|^2 E_{\alpha},
\label{energyinitialstate}
\end{equation}
and 
\begin{equation}
\sigma_0^2=\sum_{\alpha} |C^{\alpha}_{n_0}|^2 (E_{\alpha} - E_0)^2.
\label{stinitialstate}
\end{equation}

\subsection{Short Times}\label{sec:ST}
The decay of $W_{n_0}(t) $ shows different behaviors at different time scales. At extremely short times, $t \ll \sigma _0^{ - 1}$, the decay is quadratic. This is a universal behavior that does not depend on $H_0$ or $H$, but simply on $\sigma_0$. It is obtained by Taylor expanding the phase factor in Eq.~(\ref{saitorho}), 
\begin{eqnarray}
W_{n_0}(t) &&\approx \left| e^{-i E_0 t}\left[ \sum_\alpha  | C^\alpha _{n_0} |^2 - i \sum_\alpha  | C^\alpha _{n_0} |^2(E_\alpha  - E_0)t  - \frac{1}{2} \sum_\alpha  | C^\alpha _{n_0} |^2 (E_\alpha  - E_0)^2 t^2 \right] \right|^2  \nonumber\\
&& \approx  1 - \sigma _0^2 t^2 .
\label{quadratic}
\end{eqnarray}
But we are actually interested in what happens after the quadratic decay.

\subsubsection{Exponential and Gaussian decay}
\begin{figure}[ht]
\includegraphics*[scale=.368]{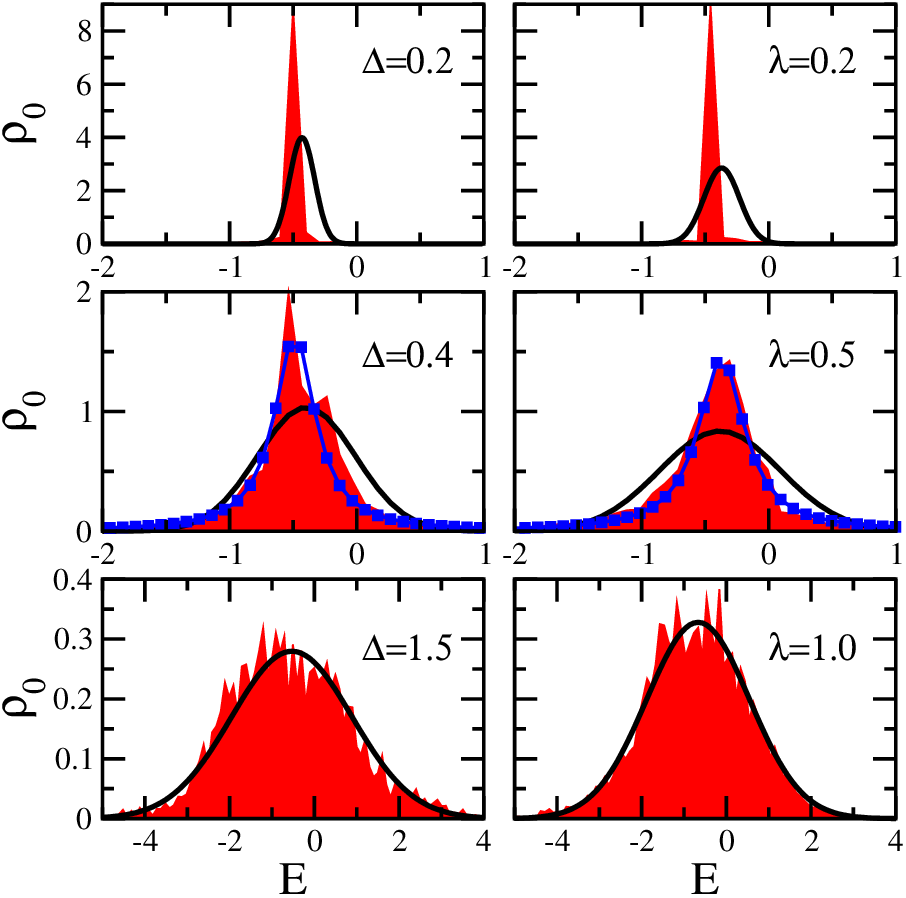}
\includegraphics*[scale=.37]{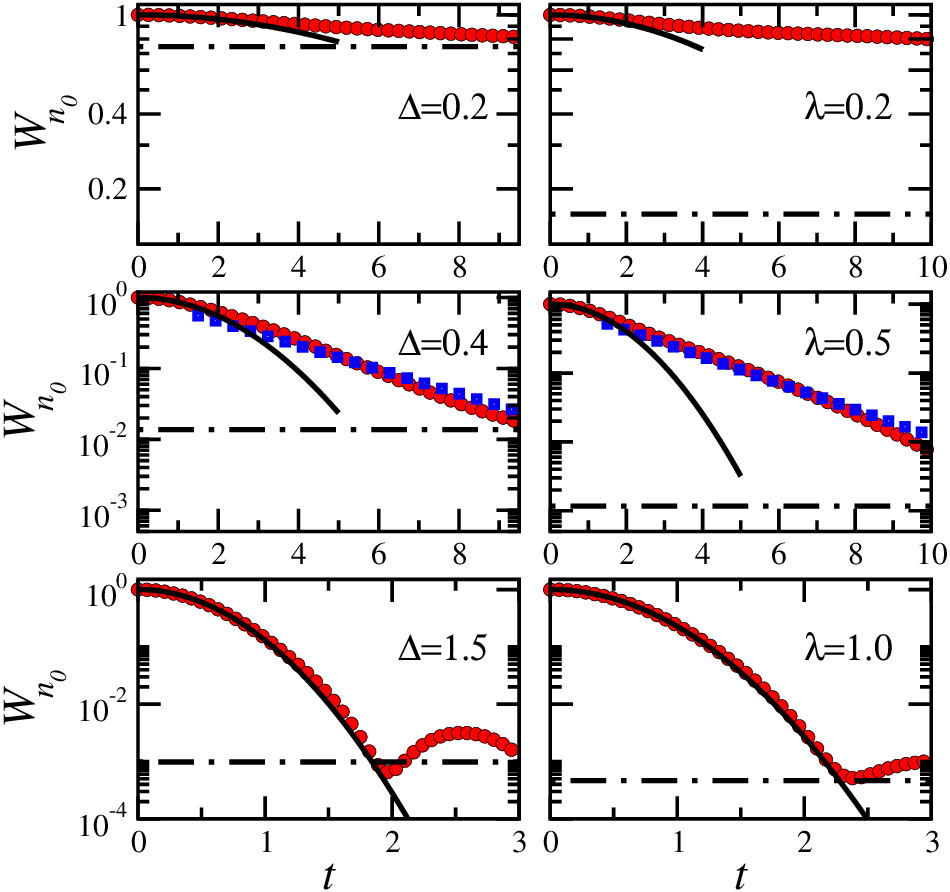}
\caption{Local density of states (two left columns) and survival probability (two right columns). First and third columns: XXZ model. Second and fourth columns: NNN model with $\Delta=0.5$. The values of $\Delta$ for the XXZ model and of $\lambda$ for the NNN model are indicated in the figure. The initial state has energy far from the edge of the spectrum; third column: $|\Psi(0)\rangle$  is an eigenstate of the XX model; fourth column: $|\Psi(0)\rangle$ is an eigenstate of the XXZ model. Blue squares: Lorentzian fit and exponential decay. Black solid line: Gaussian LDOS and Gaussian decay with $\sigma_0$ from Eq.~(\ref{stinitialstate}). Red shaded area and circles: numerical results. $L=18$, $N_{up}=6$, $h,d=0$, open chain. Horizontal dot-dashed lines indicate the saturation point [Eq.(\ref{eqSatura})].}
\label{Fig:LDOS}     
\end{figure}

After the universal quadratic behavior, the decay depends on the strength of the perturbation, which determines the shape of the LDOS.  The LDOS is close to a delta function for $g \to 0$ and its Fourier transform leads to a very slow decay of $W_{n_0}(t)$. The two left top panels of Fig.~\ref{Fig:LDOS} show LDOS for $g\to0$ and the two right top panels present the corresponding $W_{n_0}(t)$. The first and third columns of the figure are obtained for the integrable XXZ model and the initial state is an eigenstate of the XX model with $E_0$ far from the edges of the spectrum. The second and fourth columns of the figure show the results for the chaotic NNN model and the initial state is an eigenstate of the XXZ model with $E_0$ away from the borders of the spectrum.

The LDOS broadens as the strength of the perturbation increases. When the perturbation $gV$ is larger than the mean level spacing (Fermi golden rule regime), the LDOS becomes Lorentzian~\cite{Torres2014NJP,Torres2014PRE},
\begin{equation}
\rho _0(E) = \frac{1}{2\pi } \frac{\Gamma _0}{\left( E_0 - E \right)^2 + \Gamma _0^2/4} ,
\end{equation}
where $\Gamma _0$ is the width of the distribution. The Fourier transform of the Lorentzian gives the exponential decay
\begin{equation}
W_{n_0}(t) = \exp ( - \Gamma _0 t).
\end{equation}
Lorentzian LDOS and exponential decays are shown in the middle panels of Fig.~\ref{Fig:LDOS}. 

As the perturbation further increases, the LDOS widens even more and eventually becomes Gaussian~\cite{Torres2014PRA,Torres2014NJP,Torres2014PRAb,Flambaum2001a,Izrailev2006},
\begin{equation}
\rho _0(E) = \frac{1}{\sqrt {2\pi \sigma _0^2} } \exp \left[ - \frac{(E - E_0)^2}{2\sigma _0^2} \right] ,
\label{eq:GaussLDOS}
\end{equation}
This shape
reflects the Gaussian density of states. In this case, the decay of the survival probability  is Gaussian,
\begin{equation}
W_{n_0}(t) = \exp ( - \sigma _0^2 t^2).
\label{fidgaussianinterm}
\end{equation}

It is important to stress that these very fast decays of the survival probability, exponential and even Gaussian, are not exclusive to chaotic systems. As we show in Fig.~\ref{Fig:LDOS} for the XXZ model (first and third columns), fast evolutions can also happen for integrable models. The speed of the dynamics depends on the strength of the perturbation, not on the regime, integrable or chaotic, of the Hamiltonian~\cite{Santos2012PRL,Santos2012PRE,Torres2014PRA,Torres2014NJP,Torres2014PRAb,TorresKollmar2015,TorresProceed,Torres2016BJP,Torres2016Entropy,Tavora2016,Tavora2017}.

Similarities in the time evolution of integrable and chaotic models perturbed far from equilibrium can be captured also with other dynamical quantities, such as the Shannon information entropy and the von Neumann entanglement entropy. In fact, as shown in Refs.~\cite{Torres2016Entropy,Torres2017}, there is a clear parallel between the behaviors of both entropies.

We take this opportunity to mention that the equation for the evolution of observables contain the survival probability explicitly. For an observable $O$, we have
\begin{eqnarray}
O(t) &=& W_{n_0}(t) O(0)  \nonumber \\
&+& \sum_{n\neq n_0} \langle\Psi(0) | e^{i H t} |\Psi(0) \rangle O_{n_0 ,n} \langle \phi_n | e^{- i H t} |\Psi(0) \rangle \nonumber \\
&+& \sum_{n\neq n_0} \langle\Psi(0) | e^{i H t} | \phi_n \rangle O_{n,n_0 } \langle\Psi(0) | e^{- i H t} |\Psi(0) \rangle \nonumber \\
&+&\sum_{n,m\neq n_0} \langle\Psi(0) | e^{i H t} | \phi_n \rangle O_{n, m} \langle \phi_m | e^{- i H t} |\Psi(0) \rangle ,
\label{eq:TotObs}
\end{eqnarray}
where $O_{n, m} = \langle  n | O | m\rangle$ and $|\phi_n \rangle $ are the eigenstates of the initial Hamiltonian that defines the initial state. The analysis of the evolution of observables is more demanding than the study of the survival probability, since they depend on the overlaps between $|\Psi(t)\rangle$ and the other basis vectors of the initial Hamiltonian and on the details of the observables.

\subsubsection{Faster than Gaussian and Quantum Speed Limit}

\begin{figure}[ht]
\includegraphics*[scale=.48]{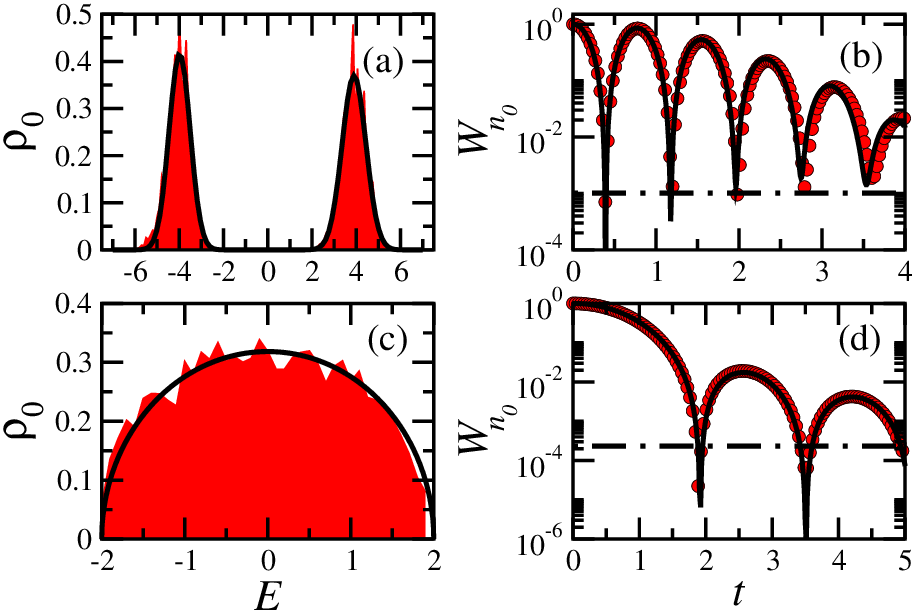}
\caption{Local density of states (left) and survival probability (right). In (a) and (b): defect model, $d=8.0$, $\Delta=0.48$, $h,\lambda=0$, $L=16$, $N_{up}=8$. Initial state in (b) is an eigenstate of the XXZ model with $E_0$ in the middle of the spectrum. In (c) and (d): Full random matrix from GOE, ${\cal D}=12870$, rescaled energies so that ${\cal E} \sim 2$. Initial state in (d) is an eigenstate of another GOE full random matrix. Red shaded areas and circles: numerical results. Black solid lines on the right: analytical expressions (\ref{fidcos}) and (\ref{eq:semicircle}). Horizontal dot-dashed lines indicate the saturation point [Eq.~(\ref{eqSatura})].}
\label{Fig:Cos}     
\end{figure}

There are scenarios where the decay of $W_{n_0}(t)$ can be even faster than Gaussian. This happens, for example, when the LDOS is bimodal (or multimodal), in which case the speed of the evolution becomes controlled by the distance between the peaks~\cite{Torres2014PRAb}. This can be achieved by preparing the system in an eigenstate of the XXZ model and evolving it with the defect model for $d\gg1$. When the amplitude of the magnetic field on the defect site is very large, the density of states and consequently also the LDOS splits in two separated Gaussian peaks, as shown in Fig.~\ref{Fig:Cos} (a). If both peaks have the same width $\sigma_G$, the Fourier transform of $\rho_0(E)$ gives
\begin{equation}
W_{n_0}(t) \simeq \cos^2(\sigma_0 t) \exp ( - \sigma_G ^2 t^2),
\label{fidcos}
\end{equation}
where $\sigma_0$ is now approximately $d/2$. 
One sees that for $t<\pi/(2 \sigma_0)$, the survival probability approaches the bound associated with the energy-time uncertainty relation, $W_{n_0}(t) \simeq \cos^2 (\sigma_0 t)$ \cite{Bhattacharyya1983,Uffink1993,GiovannettiPRA2003}. For $t>\pi/(2 \sigma_0)$ there are revivals. The envelope of the decay of these oscillations is Gaussian and controlled by $\sigma_G$. The expression in Eq.~(\ref{fidcos}) matches very well the decay of the survival probability shown in Fig.~\ref{Fig:Cos} (b).

Another example of a decay faster than Gaussian occurs for systems with random and simultaneous interactions among many particles. The extreme case is that of full random matrices. In Fig.~\ref{Fig:Cos} (c), we show the LDOS for an initial state corresponding to an eigenstate of a GOE full random matrix that is evolved with another GOE full random matrix. The LDOS has a semicircle shape~\cite{Wigner1955,Torres2014PRA,Torres2014NJP,Torres2014PRAb},  as the density of states for full random matrices [Eq.~(\ref{DOSfrm})],
\begin{equation}
\rho_0(E) = \frac{1}{\pi \sigma _0} \sqrt{ 1- \left( \frac{E}{2\sigma_0}  \right)^2} ,
\label{eq:semicircle}
\end{equation}
where $\sigma_0^2 = \int_{-\cal E}^{\cal E} \rho_0(E) E^2 dE = {\cal E}/2$. The Fourier transform of the semicircle gives the following analytical expression for the survival probability~\cite{Torres2014PRA, Torres2014NJP,Torres2016Entropy}
\begin{equation}
W_{n_0}(t) = \frac{[{\cal J}_1(2\sigma_0 t)]^2}{\sigma_0^2 t^2} ,
\label{eq:FRMdecay}
\end{equation}
where ${\cal J}_1$ is the Bessel function of the first kind. This expression agrees with the numerical results in Fig.~\ref{Fig:Cos} (d). The decay up to $t\sim{\cal E}$ is faster than Gaussian. Later, it shows oscillations that decay as a power-law $\propto t^{-3}$. Indeed, for $t\gg \sigma_0^{-1}$, Eq.~(\ref{eq:FRMdecay}) leads to
\begin{equation}
W_{n_0}(t \gg \sigma _0^{ - 1}) \to \frac{1 - \sin (4 \sigma_0t)}{2\pi \sigma_0^3 t^3}.
\label{eq:FRMasymptotic}
\end{equation}

The onset of the power-law decay for longer times, as depicted in Fig.~\ref{Fig:Cos} (d), prompts the question of what happens to the survival probability for the spin models at times longer than those shown in Fig.~\ref{Fig:LDOS}. If we wait long enough, since the studied systems are finite, the dynamics eventually saturates to the infinite-time average,
\begin{equation}
\overline{W}_{n_0}=\lim_{t\rightarrow \infty} \frac{1}{t}
\int^{t}_0 d\tau\,  F(\tau) = \sum_{\alpha} |C^{\alpha}_{n_0} |^4=IPR_{n_0},
\label{eqSatura}
\end{equation}
where $IPR_{n_0}$ is the inverse of the participation ratio of the initial state written in the energy eigenbasis. Our question is whether there is some other well defined behavior between the initial exponential or Gaussian decays and the saturation to $\overline{W}_{n_0}$. This is the subject of the next subsection.

\subsection{From Short to Long Times: Strong Perturbation} 
Since our systems are finite and relatively small, the analysis of long-time dynamics is subjected to finite size effects. To circumvent this problem, we focus now on the disordered Hamiltonian  (\ref{ham})  with $d,\lambda=0$, random uniform numbers $h_k \in [-h,h]$, $\Delta=1$, and closed boundary conditions, that is,
\begin{equation}\label{hamD}
H = \sum_{k=1}^{L} h_k S_k^z  + \sum_{k=1}^L \left( S_k^x S_{k+1}^x + S_k^y S_{k+1}^y + S_k^z S_{k+1}^z \right)\,.
\end{equation}
Hamiltonian  (\ref{hamD}) is paradigmatic in the studies of many-body localization (MBL) \cite{Santos2004,SantosEscobar2004,Santos2005loc,Dukesz2009,Nandkishore2015}. MBL refers to localization in face of the interplay between interaction and disorder. It is an extension to the Anderson localization, where interaction is absent. Without the Ising interaction, the eigenstates of the disordered noninteracting 1D system are exponentially localized in configuration space for any value of $h$. The question that has been discussed more intensely since the beginning of the millennium is whether localization may still take place when interaction is added. Although not precise, the value $h_c \approx 3.5$ for the disorder strength in (\ref{hamD}) is supposed to determine the critical point for the transition from the ergodic (chaotic) to the MBL phase.

By varying the disorder strength, we can study the survival probability at long times close to the clean integrable point ($h=0$), in the chaotic regime, in the intermediate region between ergodicity and localization, and in the MBL phase, which is another integrable point [see Fig.~\ref{Fig:indicator}]. In the chaotic regime, the eigenstates away from the border of the spectrum are highly delocalized and similar to random vectors [as in Fig.~\ref{Fig:DOS} (d)]. We refer to these states as chaotic or ergodic states, although it is important to keep in mind that ergodicity in the sense of full random matrices, where the eigenstates are random vectors, does not exist in realistic systems. As the disorder strength increases and we move from the chaotic to the MBL phase, passing through the intermediate region, the eigenstates become less spread out in space. As we discussed in \cite{Torres2017}, they remain extended in this intermediate region, but are no longer ergodic. This reduction in the level of delocalization of the eigenstates naturally slows down the dynamics.

We take as initial states, single site-basis vectors. This is equivalent to a quench, where the initial Hamiltonian is only the Ising part of Hamiltonian (\ref{hamD}) and the final Hamiltonian that dictates the evolution is the complete  $H$ (\ref{hamD}). In view of Eq.~(\ref{eq:quench}), this case corresponds to a strong perturbation. 

We perform averages over initial states and disorder realizations. This reduces finite-size effects and unveils features of the dynamics that could otherwise be hidden by sample to sample fluctuations. The average is done over $0.1 \cal{D}$ initial states with energies close to the  middle of the spectrum and over enough disorder realizations to have a total of $\sim10^5$ statistical data. The average is represented with the symbol  $<.>$.  We choose initial states with energy close to the middle of the spectrum ($E_0\approx0$), because there localization is more difficult, due to the large concentration of energy levels. If localization occurs at $E_0\approx0$, then it is certain to have taken place at other regions of the spectrum.

Figure~\ref{fig:disorder} depicts the time evolution of the averaged survival probability, $\langle W_{n_0}(t)\rangle$, from very short to very long times. The disorder strength $h$ ranges from $h=0.2$ (chaotic regime) to $h=4.0$ where the system is likely already in the MBL phase. 

According to Eq.~(\ref{quadratic}), the dynamics at very short times ($t\ll \sigma_0^{-1}$) depends only on $\sigma_0$. If we write the Hamiltonian matrix in the site-basis (denoted by $|\phi_n \rangle$), we can show that 
\begin{equation}
\sigma_0=\sum_{\alpha} |C^{\alpha}_{n_0}|^2 (E_{\alpha} - E_0)^2 = \sqrt{\sum_{n\neq n_0}|\langle\phi_n|H|\phi_{n_0}\rangle|^2} \,,
\label{stinitialstate2}
\end{equation}
where $|\phi_{n_0}\rangle = |\Psi(0)\rangle$.
In the site-basis, the disorder appears only in the diagonal elements of the Hamiltonian matrix. Thus, the dynamics at very short times is completely independent of the presence of disorder.

The subsequent evolution is purely Gaussian, as expected from the Gaussian envelope of the LDOS [see Fig.~\ref{Fig03LDOS}]. The evolution in this time scale agrees very well with the analytical expression 
$\langle W_{n_0}(t)\rangle=\exp(-\sigma_0^2 t^2)$ discussed in Eq.~(\ref{fidgaussianinterm}). This is illustrated with circles in Fig.~\ref{fig:disorder}. When $h$ becomes large, the time interval of the Gaussian decay shortens, and possibly only the quadratic part of the decay persists. 

\subsubsection{Power-law decays}
\begin{figure}[ht]
\includegraphics*[scale=0.4]{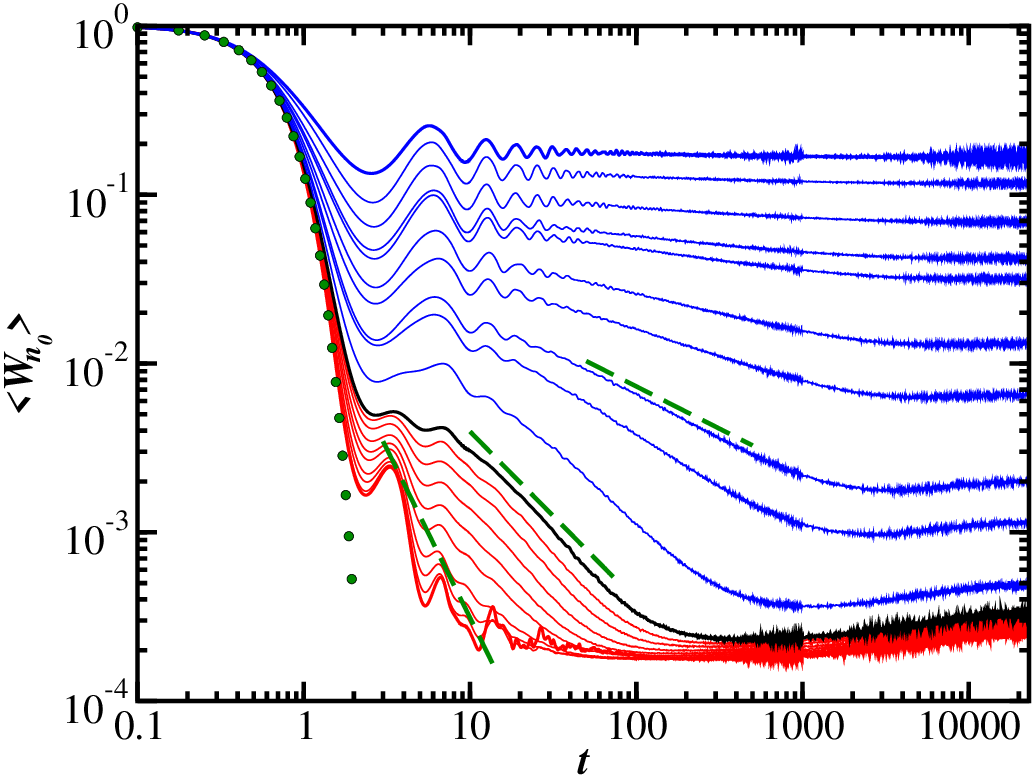}
\caption{Survival probability. From bottom to top, $h=0.2, 0.3, \ldots 0.9$, $h=0.95, 1, 1.25, \ldots 3$, $h=3.5, 4$. Circles: analytical Gaussian decay $\langle W_{n_0}(t)\rangle= \exp(-\sigma_0^2 t^2)$. Dashed lines are, from bottom to top, $\gamma = 2$,  $\gamma = 1$, and  $\gamma = 0.5$ for $h=0.2$, $h=1.0$ and $h=1.75$, respectively. Averages over $10^5$ data of disorder realizations and initial states with $E_0 \sim 0$; $L=16$, $N_{up}=8$.}
\label{fig:disorder}  
\end{figure}

After the fast Gaussian evolution, oscillations emerge. These are not fluctuations that could be reduced with a large number realizations, as those after equilibrium. These oscillations may in fact belong to the power-law decays that become evident  in Fig.~\ref{fig:disorder} for $t>2$.

The power-law exponent $\gamma$ in $\langle W_{n_0}(t)\rangle\propto t^{-\gamma}$ depends on the disorder strength $h$. The two different colors in Fig.~\ref{fig:disorder} (red and blue curves) indicate two different causes of the power-law decay, as we discuss next.

\begin{figure}[ht]
\includegraphics*[width=4.5in]{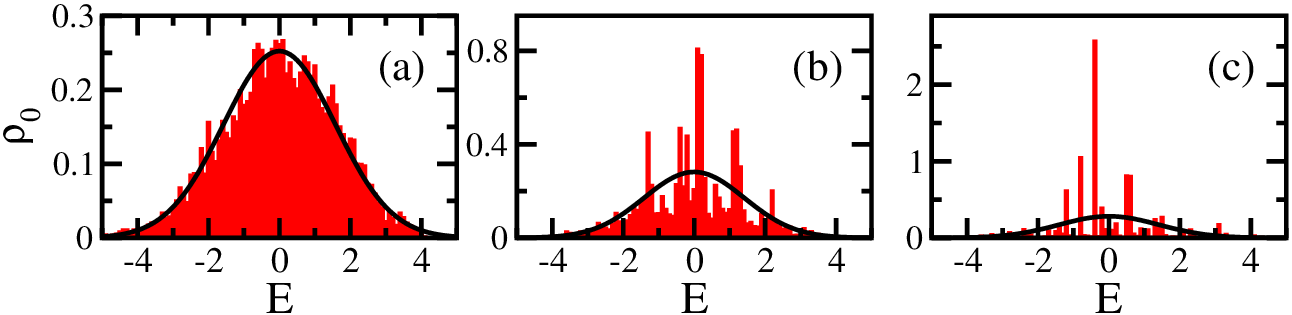}
\caption{Local density of states for a single disorder realization; $h=0.5$ (a), $h=1.5$ (b) and $h=2.7$ (c);  $L=16$, $N_{up}=8$. The envelopes (solid lines) of the distributions  are Gaussians with center $E_{0}$ [Eq.~(\ref{energyinitialstate})] and width $\sigma_{0}$ [Eq.~(\ref{stinitialstate2})], which is consistent with the situation of strong perturbation.}
\label{Fig03LDOS}  
\end{figure}

\paragraph{Chaotic eigenstates}
The red curves in the bottom of Fig.~\ref{fig:disorder} are associated with the results for the system in the chaotic domain. According to Fig.~\ref{Fig:indicator}, this occurs for $0.1<h<1$. In this region, the LDOS is well filled as seen in Fig. \ref{Fig03LDOS} (a) for $h=0.5$. The analysis of the participation ratio of the initial state confirms ergodicity, $\langle \text{PR}_{0} \rangle \propto {\cal D}$. In this region, we expect $\gamma$ to be close to 2, as is indeed obtained with the curve for $h=0.2$ that is shown in the bottom of Fig.~\ref{fig:disorder} together with the dashed line that represents $\langle W_{n_0}(t)\rangle \propto t^{-2}$. The exponent $\gamma=2$ is caused by the so-called Khalfin effect. It refers to the emergence of the power-law decay of the survival probability due to the unavoidable presence of bounds in the spectrum~\cite{Khalfin1958,MugaBook,Urbanowski2009,Campo2016}. The phenomenon has been extensively studied for continuous systems. We have argued that similar analyses can be extended to the discrete spectra of finite lattice many-body quantum systems when the LDOS is ergodically filled~\cite{Tavora2016,Tavora2017}. 


Notice, however, that as $h$ increases above $0.2$ up to $1$, $\gamma$ decreases from $2$ and approaches $1$, as seen in  Fig.~\ref{fig:disorder}. These intermediate values, $1\leq \gamma <2$, are probably caused by a competition between the effects of energy bounds and minor correlations between the eigenstates. 

\paragraph{Multifractal eigenstates}

The black curve in Fig.~\ref{fig:disorder} marks the borderline between the chaotic region (red) and the intermediate region (blue). In the latter, the eigenstates become multifractal. Multifractality implies that the sums of the moments $M$ of the components $|C_n^{\alpha}|^2$ of the eigenstates $|\psi_{\alpha} \rangle = \sum_n C_n^{\alpha} |\phi_n \rangle $ written in the site-basis $|\phi_n \rangle $ show multifractal scaling with the dimension of the Hilbert space ${\cal D}$, that is
\begin{equation}
\langle M_q \rangle= \sum_n  |C_n^{\alpha}|^{2q} \sim {\cal D}^{-(q-1) D_q},
\end{equation}
where $D_q$ is the fractal dimension. Multifractality occurs when $D_q$ depends nonlinearly on $q$, instead of being a constant, as in the metallic ($D_q=1$) or in the insulating ($D_q=0$) phase. Most of our studies have concentrated on the second moment $M_2$ for the eigenstates written in the site-basis and for the initial states (which are site-basis vectors) written in the energy eigenbasis. Our focus has therefore been on $D_2$. The second moment is nothing but the participation ratio, $PR^{(\alpha)}$ for the eigenstates and $PR_0$ for the initial states. We calculated $D_1$ in \cite{Torres2017} and other $q$'s have been recently studied as well~\cite{SerbynARXIV}.

Our scaling analyses for $PR^{(\alpha)}$ and $PR_0$ suggest that both lead to the same value of $D_2$. This value is $\sim 1$ in the chaotic region and $<1$ in the intermediate region. The intermediate region is therefore characterized by eigenstates that are not yet localized, but are not chaotic either. In this region, $D_2$ decreases as $h$ increases. The fractality of the states also leads to the sparsity of the LDOS, as seen in Fig.~\ref{Fig03LDOS} (c) and Fig.~\ref{Fig03LDOS} (d).

We got excited when we realized that in the intermediate region, the value of $D_2$ coincides with exponent of the power-law decay $\gamma$, that is $\langle W_{n_0}(t)\rangle\propto t^{-D_2}$. This agreement is better understood if one writes the survival probability in terms of the correlation function ${\cal C}(E)= \sum _{\alpha_1 , \alpha_2} |C^{(\alpha_1)}_{n_0} |^2 | C^{(\alpha_2)}_{n_0} |^2 \delta( E - E_{\alpha_1} + E_{\alpha_2}  ) $ as follows
\begin{equation}
W_{n_0}(t)  = \int_{ - \infty }^\infty  dE e^{iE t} {\cal C}(E ) .
\label{FtfromComega}
\end{equation}
The long-time behavior of $W_{n_0}(t)$ is determined by small $E$. A power-law decay with exponent $D_2$ emerges for long $t$ when~\cite{Chalker1988,Chalker1990,Ketzmerick1992,Huckestein1994,Huckestein1999,Cuevas2007,Kravtsov2011} 
\begin{equation}
{\cal C}(E   \to 0) \propto E^{D_2  - 1}.
\end{equation}
This is analogous to what has been found in noninteracting disordered systems at least as early as in the studies by Chalker~\cite{Chalker1988,Chalker1990}.

During the time interval where the power-law decays with $\gamma=D_2<1$ are seen for the survival probability, we observe also a logarithmic growth of the Shannon entropy and entanglement entropy controlled by the same fractal dimension as $S\sim A + D_2 \ln t$ ($A$ is a constant).

\subsubsection{Correlation hole}
After the power-law decay,  there is still one more interesting feature in the decay of the survival probability  before it finally saturates to $\overline{W}_{n_0}$. The survival probability may fall below the saturation value and then raise to $\overline{W}_{n_0}$. This dip is known as correlation hole~\cite{Leviandier1986,Guhr1990,Alhassid1992,Gorin2002}. It is an explicit dynamical manifestation of level repulsion; it only appears in nonintegrable finite systems~\cite{Torres2017,Torres2017Philo}. Thus, by studying the evolution of the survival probability at long times, we gain information about level statistics. The correlation hole is visible in Fig.~\ref{fig:disorder}, being clearly deeper in the chaotic region. As the disorder strength increases and the system approaches the MBL phase, the hole fades away and eventually disappears.

We can use the depth of the correlation hole to quantify how close or far the system is to the chaotic region. 
To measure the depth, we compute
\begin{equation}
\kappa = \frac{\overline{W}_{n_0} - \langle W_{n_0}^{min}\rangle}{\overline{W}_{n_0}} .
\end{equation}
In full random matrices from GOE, $W_{n_0}^{min} \sim 2/{\cal D}$  \cite{Alhassid1992} and $\overline{W}_{n_0} \sim 3/{\cal D}$, so the maximum value that $\kappa$ can have is $1/3$.

\begin{figure}[ht]
\includegraphics*[width=4.5cm]{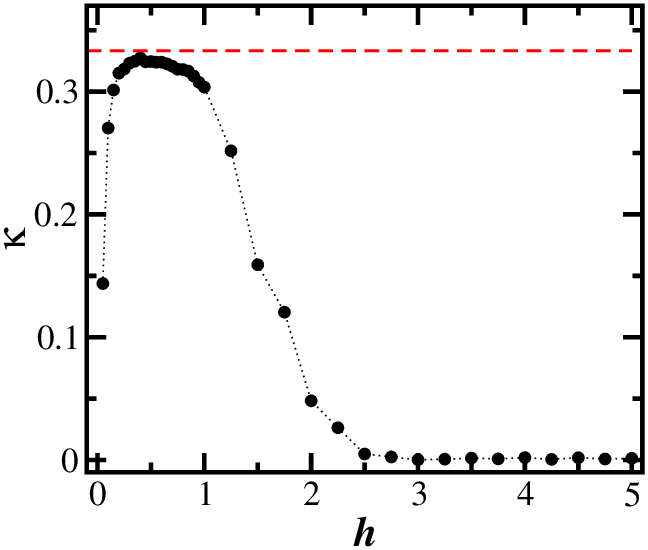}
\caption{\label{Fig:CH}
Depth $\kappa$ of the correlation hole {\em vs.} disorder strength $h$. Dashed line corresponds to the result for full random matrices, $\kappa^{FRM}=1/3$. $L=16$, $N_{up}=8$.}
\end{figure}

In Fig.~\ref{Fig:CH}, we show  $\kappa$ as a function of the disorder strength. It approaches the maximum value $1/3$ in the chaotic region. It decreases for small $h$, since the system gets closer to the integrable clean point, and for large $h$, as the system approaches localization. Similarly to $D_2$, $\kappa$ is another alternative to detect the transition from chaos to spatial localization. But notice that $\kappa$ can in fact detect and integrable-chaos transition.

The correlation hole is seen also in observables, such as the spin density imbalance. Using full random matrices we were able to find exact analytical expressions for the survival probability and for the imbalance from $t=0$ to saturation~\cite{Torres2017ARXIV}. These expressions helped us to propose functions that matched very well the entire evolution of the survival probability of spin systems deep in the chaotic region, including the correlation hole, and that captured very well different behaviors of the imbalance at different time scales~\cite{Torres2017ARXIV}.

\section{Conclusions}
We close this chapter with a very brief discussion about future plans. In addition to the immediate goal of extending the level of details that we have obtained for the survival probability to other physical observables, we intend to explore how our studies are affected by couplings with an environment. As we have shown, the dynamics of isolated many-body quantum systems depends on several factors, such as the energy of the initial state, the strength of the perturbation that takes the system out of equilibrium, the regime of the Hamiltonian (whether integrable or chaotic), the presence of disorder, the strength of the interactions, and the number of particles that interact simultaneously. Despite these many factors and the different behaviors at different time scales, we have been able to extract general features. What should happen to our picture when external interactions are also included? What will follow from the interplay between internal and external interactions? Which will be the dominant elements controlling the dynamics and can they have different roles at different time scales?

\begin{acknowledgments}
LFS was supported by the NSF grant No. DMR-1603418. EJTH acknowledges funding from PRODEP-SEP and Proyectos VIEP-BUAP, Mexico.
\end{acknowledgments}

\end{document}